\documentclass[runningheads]{llncs}
\usepackage[T1]{fontenc}
\usepackage{graphicx}
\usepackage{url}
\usepackage{etoolbox}
\usepackage{float}
\apptocmd\thebibliography{\small}{}{}

\begin{document}

\title{Culturally-Aware AI for Cross-Boundary Community Learning: Undergraduate Innovation at the Intersection of Computation and Design}

\titlerunning{Culturally-Aware AI for Cross-Boundary Community Learning}

\author{
Jiaojiao Zhao\inst{1} \and
Weisheng Zhang\inst{1} \and
Jiawen Cai\inst{1} \and
Haibin Gao\inst{2} \and
Luyao Zhang\inst{1}\thanks{\footnotesize The corresponding author: Luyao Zhang (email:lz183@duke.edu), Digital Innovation Research Center and Social Science Division, Duke Kunshan University. Address: Duke Avenue No.8, Kunshan, Suzhou, Jiangsu, China, 215316. \textbf{Acknowledgments}: This work was supported by the Community-Based Learning (CBL) Grant from the Office of Student Experience and the Center for the Study of Contemporary China (CSCC), Duke Kunshan University. Jiaojiao Zhao, Weisheng Zhang, and Dr. Luyao Zhang gratefully acknowledge the enriching exchanges with participants at the Third Cross-Strait Hong Kong and Macao Service-Learning Student Conference at The Hong Kong Polytechnic University (2026), whose perspectives on service leadership and community-engaged learning—especially those shared by Prof. Daniel Shek and the representatives from Tin Ka Ping Foundation—deeply informed this work. An earlier version of this work was presented at the 8th International Workshop on Culturally-Aware Tutoring Systems (CATS 2026), co-located with AIED 2026 in Seoul. The authors thank the workshop organizers—Ivon Arroyo, Emmanuel G. Blanchard, Christine Kwon, and Maria Mercedes T. Rodrigo—and workshop participants for constructive exchanges that sharpened the paper's conceptualization of culture and its discussion of culturally-aware design across contexts.}
}

\authorrunning{J. Zhao, W. Zhang, J. Cai, H. Gao, L. Zhang}

\institute{
Duke Kunshan University, Suzhou, China
\and
Zhouzhuang Mystery of Life Museum, Suzhou, China
}

\maketitle

\begin{abstract}
Research on artificial intelligence in education (AIED) is rapidly expanding, yet technical progress often lacks human-centered grounding and adequate attention to cultural context. Community-Based Learning, a pedagogy rooted in social work, remains underrepresented in AIED research, particularly within Asia-Pacific contexts. This paper reports on cross-boundary Community-Based Learning where undergraduate students develop AI-enabled solutions for cultural heritage preservation and sustainable development. We examine how community-engaged computing operationalizes culturally aware, human-centered AIED through participatory elicitation of cultural knowledge, bilingual representation, and stakeholder validation across education, technology, and culture. We contribute a collaborative framework for culturally aware AIED designed to support multi-stakeholder collaboration and widen participation by bridging social work and computational science.

\end{abstract}

\section{Introduction}\label{sec:intro}
\begin{figure}[!t]
\centering
\includegraphics[width=0.95\linewidth]{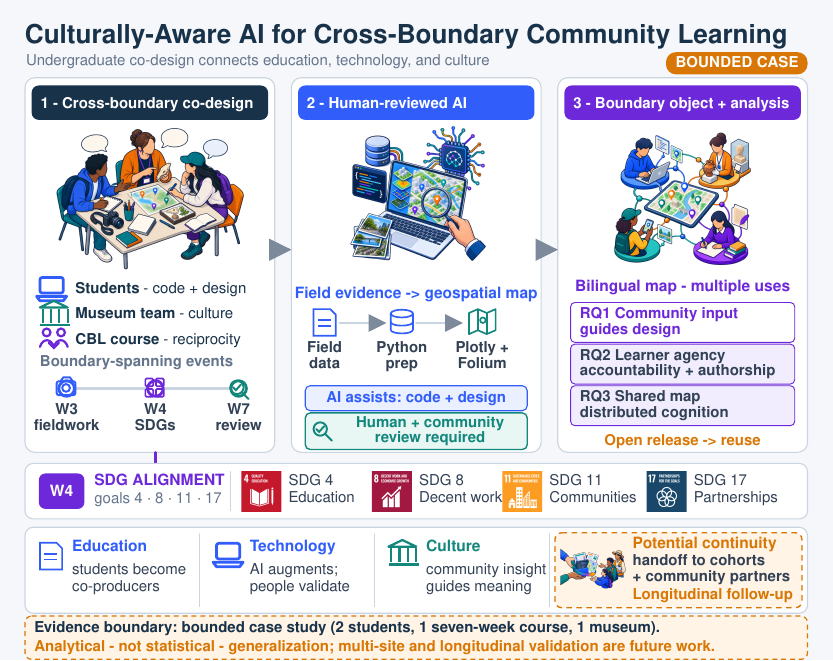}
\caption{\textbf{Teaser.} Overview of the community-based learning (CBL) project context and the AI-enabled cultural heritage prototype developed with community partners.}
\label{fig:teaser}
\end{figure}

As shown in Figure~\ref{fig:teaser}, we study a cross-boundary community-based learning setting where undergraduate students develop AI-enabled solutions for cultural heritage preservation and sustainable development.

Recent advances in artificial intelligence in education (AIED), driven by
generative AI and large language models, have intensified scholarly attention
to human--AI collaboration and ethical design \cite{Feng2025MappingAIED}. Yet these developments frequently proceed
without adequate stakeholder involvement, raising concerns regarding privacy,
agency, and alignment \cite{Alfredo2024HCLA}. This study addresses an
underrepresented dimension: community-engaged pedagogies that widen
participation beyond classroom boundaries in culturally diverse Asia-Pacific
contexts. Recent bibliometric analyses reveal significant geographical
imbalances in AIED research, with scholarly production concentrated in North
America and Europe while Asia-Pacific ecosystems remain underrepresented
\cite{10.1145/3746469.3746555}. This imbalance also raises a
generalizability concern: evidence drawn from narrow Western populations
should not be assumed to represent learners universally
\cite{henrich2010weird}. The CATS 2026 agenda identifies three linked
challenges for culturally-aware AIED: selecting relevant cultural features,
testing transfer across contexts, and embedding cultural responsiveness in
learner--AI interaction \cite{arroyo2026cats}. Following the
multicultural-education tradition articulated by Pusch, we treat culture as
a shared but variable system of meanings, values, practices, and artifacts,
rather than as a fixed national label \cite{pusch1979multicultural}.

We focus on Community-Based Learning
\cite{ridwan2025beyond,butcher2025community}---a pedagogy leveraging digital
technologies to bridge academic and community contexts, fulfilling the ``Third
Mission'' of universities to generate social impact \cite{butcher2025community}.
Conventionally anchored in social work, Community-Based Learning functions here
as a cross-boundary mechanism: importing social work's engagement methodologies
into computational training and creating sustainable university-community
partnerships. This reconfigures AIED by treating students as knowledge
producers---final projects are released as open-source software, circulating
innovations within the AI-for-social-good ecosystem.

Central is collective intelligence---the capacity of groups to
solve problems exceeding individual capabilities through collaboration and
distributed cognition \cite{Casebourne2025CI}. By releasing projects as
open-source and including students as co-authors, this pedagogy structures
learning to generate collective outcomes within distributed knowledge
ecosystems \cite{Casebourne2025CI}. We frame culturally-aware AIED across
three dimensions: education, technology, and culture.
We ask:

\textbf{RQ1.} How do students incorporate community stakeholder input when
developing AI-enabled prototypes for cultural benefit and United Nations' Sustainable Development Goals (SDGs)?

\textbf{RQ2.} What forms of learner agency and accountability emerge when
students design and present community-based AI prototypes?

\textbf{RQ3.} How can collective intelligence evaluate community-engaged AIED
outcomes beyond individual performance?

\section{Background and Method}\label{sec:method}

\subsection{Educational Context and Participants}\label{subsec:context}

Recent bibliometric analyses reveal that Asia-Pacific research ecosystems remain systematically underrepresented in AIED \cite{10.1145/3746469.3746555,blanchard2015socio}. This study reports a bounded case study \cite{case2024} of two undergraduate students---the first two authors---who teamed up in INFOSCI 301 \textit{Data Visualization and Information Aesthetics}, a 7-week session (Fall 2025) at Duke Kunshan University. The course serves the Computation and Design major, an interdisciplinary undergraduate curriculum leading to dual degrees awarded by Duke University (US) and Duke Kunshan University (China). The cap enrollment for each 7-week session is $N=18$, spanning students from three divisions: Arts and Humanities, Social Science, and Natural and Applied Sciences. Our aim is analytical generalization---generating theoretically-grounded propositions about culturally-aware AIED design \cite{Pratt2024}---rather than statistical generalization.

The course integrates technical training in data processing, geospatial visualization (Plotly, Folium), and software engineering with Community-Based Learning methodologies. UNESCO identifies Community-Based Learning as a critical yet underutilized framework for higher education's Third Mission \cite{butcher2025community}.
\begin{figure}[!htbp]
\centering
\includegraphics[width=0.88\linewidth]{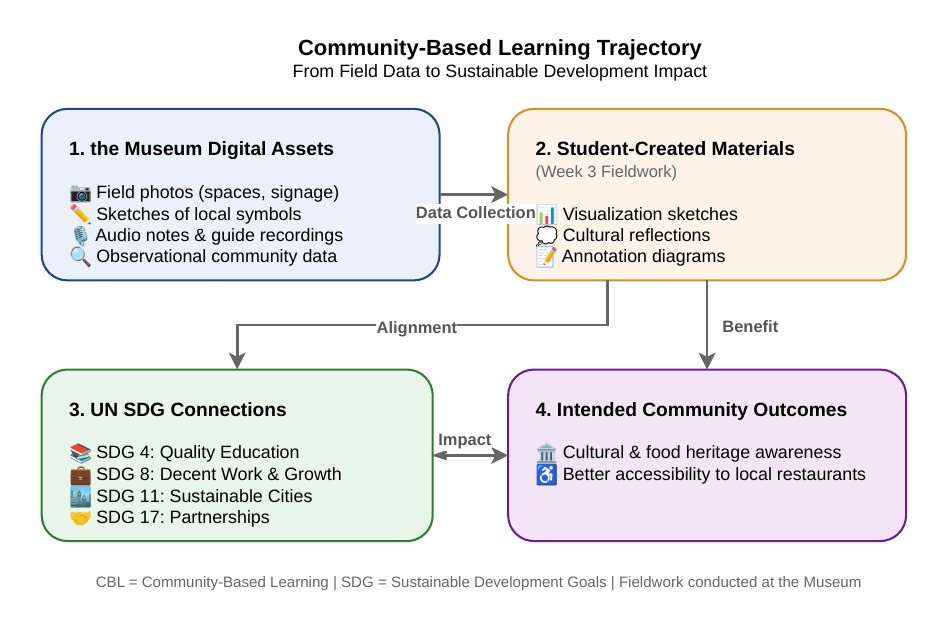}
\caption{\textbf{CBL Trajectory} from Cultural Field Data to Sustainable Development Impact. Four stages: (1)~Museum Digital Assets from Week~3 fieldwork; (2)~Student-Created Materials; (3)~UN SDG Connections via Week~4 alignment; (4)~Intended Community Outcomes validated through Week~7 participatory feedback.}
\label{fig:concept_flow}
\end{figure}

\textbf{Museum Partnership.} We partnered with the \textit{Zhouzhuang Mystery of Life Museum}, a regional natural history museum in Kunshan, Jiangsu Province, specializing in local ecological and cultural heritage. The Museum was selected through the university's Community-Based Learning Grant program based on three criteria: (1) geographic proximity enabling repeated site visits; (2) alignment between the Museum's spatial storytelling mission and the course's visualization learning objectives; and (3) the Museum's expressed need for digital extensions of its physical exhibitions---a common challenge for small regional cultural institutions \cite{pandya2025transformative}. Three museum educators participated as bilingual (Mandarin/English) community partners.

\textbf{Cultural Heritage Focus.} The two students identified local culinary culture as the design focus during Week 2 fieldwork for three reasons. First, the Museum's exhibitions emphasize natural history and geological heritage, while the surrounding town's living cultural fabric---its restaurants, local dishes, and food traditions---offered a rich, undocumented dataset that students could meaningfully contribute to. Second, restaurants function as distributed, accessible cultural nodes extending the Museum's spatial storytelling mission into residents' daily lives. Third, the bilingual (Chinese/English) map format supports place-based navigation for both local residents and international visitors \cite{chaouch2026cultural}.

\subsection{Participatory Co-Design and Technical Architecture}\label{sec:codesign}

We employ participatory action research \cite{Pratt2024} within case study methodology \cite{case2024}, incorporating three \textit{boundary-spanning events}---structured occasions traversing institutional interfaces---following Star and Griesemer's \cite{Star1989} concept of \textit{boundary objects}, artifacts enabling coordinated action across social worlds while supporting divergent interpretations. Figure~\ref{fig:concept_flow} depicts the four-stage CBL trajectory linking museum field data to sustainable community impact. Our work situates within broader efforts to embed culturally-adaptive mechanisms into intelligent educational technologies \cite{blanchard2010infusing} and recent investigations of cultural intelligence in LLM-based tools \cite{blanchard2024cultural}. Our case complements MAUOC's formal modeling of cultural concepts with participatory elicitation: community partners help determine relevant categories, bilingual labels, and validation criteria \cite{blanchard2014mauoc}. Bennett's developmental model further motivates movement from recognizing cultural difference toward acceptance and adaptation \cite{bennett1986intercultural}; without a pre/post intercultural-sensitivity measure, however, we treat the activities as opportunities for such development rather than evidence of stage progression.

\textit{Event 1 (Week~3): Field Co-Design.} Students visited the Museum to establish co-design dialogue on spatial storytelling \cite{ridwan2025beyond}, generating Museum Digital Assets and Student-Created Materials. They explored extending spatial storytelling into urban food culture---treating restaurants as elements of a distributed ``open-air exhibition'' for place-based navigation.

\textit{Event 2 (Week~4): Collaborative Alignment.} A mock symposium where students synthesized fieldwork into proposals aligning with four SDGs: SDG~4 (Quality Education), SDG~8 (Decent Work), SDG~11 (Sustainable Cities), and SDG~17 (Partnerships) \cite{butcher2025community,ridwan2025beyond}.

\textit{Event 3 (Week~7): Mutual Validation.} Students presented prototypes to community partners, staff, and peers, enacting mutuality where validation criteria are negotiated between stakeholders \cite{pandya2025transformative,gulikers2025transdisciplinary}. Following validation, the team engaged in co-dissemination: (a)~MIT-licensed open-source release as digital public goods \cite{pandya2025transformative}; and (b)~student co-authors drafting manuscript sections, operationalizing \textit{epistemic equity}---the fair distribution of knowledge-production roles across institutional hierarchies \cite{pandya2025transformative}.

\begin{figure}[!t]
\centering
\includegraphics[width=0.78\linewidth]{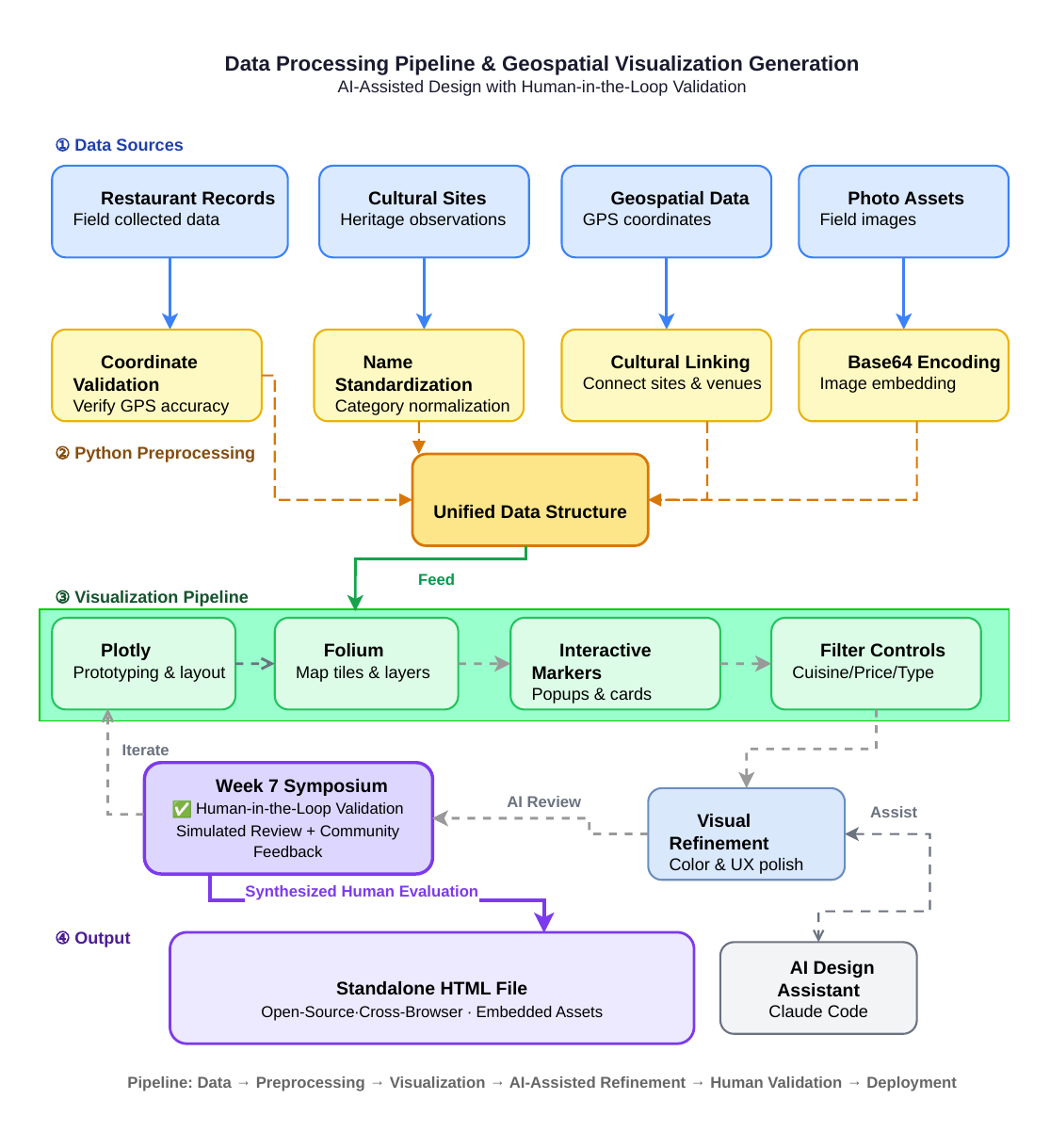}
\caption{\textbf{Data Processing Pipeline and Cultural Visualization Workflow.} End-to-end workflow from heterogeneous raw data ingestion through Python preprocessing to unified data structure, followed by Plotly/Folium visualization with AI-assisted refinement and human-in-the-loop validation, culminating in standalone HTML deployment.}
\label{fig:data-flow}
\end{figure}

The technical pipeline (Figure~\ref{fig:data-flow}) integrates heterogeneous data sources through Python preprocessing---coordinate validation, name standardization, cultural annotation linking, and Base64 encoding---into a unified data structure. Visualization proceeds through Plotly prototyping and Folium map rendering with interactive markers and filter controls. Claude Code served as an AI-assisted design partner for four functions: (1)~code generation and debugging; (2)~synthetic interface review evaluating color contrast, bilingual label clarity, and cultural sensitivity; (3)~data preprocessing assistance; and (4)~documentation drafting. All AI-generated outputs underwent mandatory human-in-the-loop review. This protocol operationalizes human-centered AIED principles \cite{Alfredo2024HCLA} by positioning AI as augmentative rather than substitutive \cite{baker2016stupid}. The final output is an interactive
website with replicable data and source code released as open access on a public GitHub repository under MIT
license at \url{https://github.com/Rising-Stars-by-Sunshine/DiscoverKunshan}.

\section{Results and Discussion}\label{sec:results}

\subsection{RQ1: Integrating Community Stakeholders in Culturally-Aware AI Design}

\begin{figure}[!t]
\centering
\includegraphics[width=0.85\linewidth]{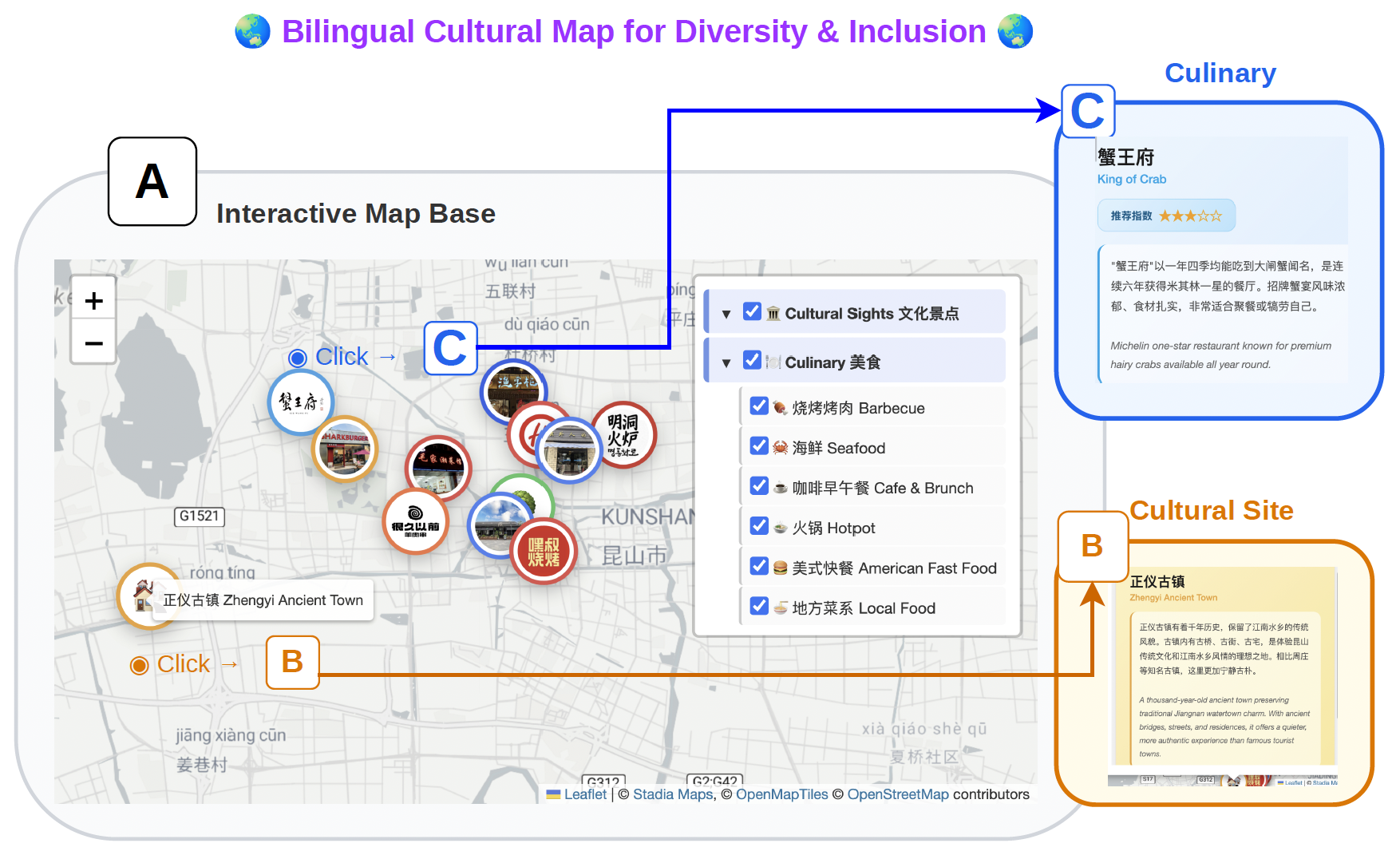}
\caption{\textbf{Bilingual Cultural Map Interface.} (A)~Interactive map
base with categorized location markers for culinary and cultural sites;
(B--C)~Bilingual pop-up panels for Cultural Sites and Culinary spots.}
\label{fig:map_teaser}
\end{figure}

Co-design dialogue with Museum educators revealed an unmet need: while the Museum possessed expertise in physical exhibitions, it lacked digital infrastructure to extend spatial storytelling into urban cultural contexts---a pattern consistent with broader challenges small regional cultural institutions face worldwide \cite{blanchard2015socio}. Students designed the \textit{Bilingual Cultural Map Interface}
(Figure~\ref{fig:map_teaser})---treating urban food spaces and heritage
sites as a distributed ``open-air exhibition'' extending the Museum's mission.

Students implemented bilingual (Chinese/English) interfaces and
high-contrast visual encodings for inclusive science communication,
aligning with Community-Based Learning principles of cultural relevance
\cite{ridwan2025beyond} and supporting SDG~4 (equitable access) and SDG~11
(inclusive cultural participation). The Week~4 and Week~7 events
established validation extending beyond the Museum to local economic
viability (SDG~8) and multi-stakeholder partnerships (SDG~17)
\cite{gulikers2025transdisciplinary}.

\subsection{RQ2: Student Agency and Accountability}

The MIT-licensed release enacted reciprocity as digital public goods
\cite{pandya2025transformative}. Repository logs document iterative
refinements and ethical considerations. The validation process enacted
mutuality---criteria negotiated between stakeholders
\cite{gulikers2025transdisciplinary}. Student reflections noted presenting
to community partners created ``pressure to ensure the project actually
worked for the community, not just for the grade'' \cite{pandya2025transformative}.
Student co-authors drafted sections analyzing their own learning,
transitioning from ``data sources'' to knowledge producers. This
arrangement embodies collective intelligence by distributing scholarly
authority across institutional hierarchies \cite{Casebourne2025CI},
enacting \textit{epistemic equity}---the fair distribution of knowledge-production roles and scholarly authority across institutional boundaries---and shifting from deficit-based to asset-based approaches
acknowledging the capabilities of all participants
\cite{pandya2025transformative}.

\subsection{RQ3: Collective Intelligence in Cross-Boundary Learning}

Applying collective intelligence \cite{Casebourne2025CI} reveals three
emergent properties.

\textbf{Interdisciplinary Knowledge Synthesis.} The AI-enabled Community-Based
Learning instantiation advances from transactional ``helpers and recipients''
toward reciprocal knowledge partnerships \cite{yahaya2025mapping},
addressing the gap where computation and design training privileges problem-solving over
ethical reflection \cite{lozano2026active}. Students bridged computational
and social work methodologies, translating technical choices into commitments
to human dignity and intercultural dialogue.

\textbf{Philanthropic-Academic-Community Triads.} At the Third Cross-Strait Hong Kong and Macao Service-Learning Student Conference at The Hong Kong Polytechnic University (2026), student co-authors presented the Bilingual Cultural Map Interface to an audience comprising student peers from 38 universities, social work faculty, and philanthropic foundation representatives. The presentation documented how open-source geospatial infrastructure enables longitudinal community benefit: the codebase remains publicly accessible beyond the semester, allowing subsequent cohorts and community partners to extend the cultural dataset rather than rebuilding from scratch. This mechanism of ``paying it forward'' through shared technical resources transforms what would otherwise be a disposable course assignment into persistent civic infrastructure \cite{pandya2025transformative}.

\textbf{Distributed Cognition.} The Bilingual Cultural Map Interface functions
as a \textit{boundary object}---an artifact that enables coordinated action across social worlds while supporting divergent interpretations \cite{Star1989}: students view it as code architecture; museum educators as curriculum infrastructure; end-users as cultural heritage for place-based learning. It accommodates competing priorities across sectors without requiring semantic consensus, sustaining SDG-focused partnerships.

\section{Implications}\label{sec:implications}

Our cross-boundary Community-Based Learning framework validates a reciprocal dissolution: CBL methodologies address computation and design education's ethical reflection gaps, while computational infrastructure addresses CBL's sustainability challenges \cite{butcher2025community,gulikers2025transdisciplinary,pandya2025transformative}. The human-in-the-loop validation protocol institutionalizes cultural accountability through mandatory community partner review of all AI outputs \cite{baker2016stupid,Alfredo2024HCLA}. Future work should scale the co-design model to underrepresented cultural contexts, assess longitudinal community adoption, and comparatively evaluate AI-assisted design tools against human-only baselines. As a single-site pilot, generalizability requires multi-site validation. By aligning technical education with cultural heritage preservation, this framework opens future research at the intersection of AI-assisted pedagogy, community-engaged design, and sustainable development across diverse socioeconomic and linguistic contexts.
%
\bibliographystyle{splncs04}
\bibliography{mybibliography}

%
%
\newpage
\appendix

\section{Extended Analysis: Collective Intelligence and Future Directions}
\label{app:extended}

This appendix provides the extended analysis of collective intelligence
and future research trajectories. Section~\ref{app:cross} elaborates on
cross-boundary knowledge exchange. Section~\ref{app:boundary} extends
the boundary object analysis. Section~\ref{app:rq3-full} presents the
full RQ3 analysis that was abbreviated in the main text.

\subsection{Cross-Boundary Knowledge Exchange}
\label{app:cross}

The synthesis of computational engineering and social work methodologies
provided the conceptual framework for negotiating ethical dimensions absent
in purely technical curricula. By importing community-engagement practices
from social work into software development, students acquired the vocabulary
to translate technical choices into tangible commitments to human dignity
and intercultural dialogue.

At the Third Cross-Strait Hong Kong and Macao Service-Learning Student
Conference at The Hong Kong Polytechnic University (2026), student
co-authors presented their work to student peers from 38 universities,
social work faculty, and philanthropic foundation representatives. For
computational education, importing Community-Based Learning methodologies
positions technical work as sustainable civic infrastructure that outlasts
semester timelines \cite{lozano2026active}. Reciprocally, for service-learning
scholarship, which documents persistent sustainability challenges and short
funding cycles \cite{yahaya2025mapping,ridwan2025beyond}, the AI-enabled
digital technology---geospatial visualization, open-source infrastructure,
human-AI collaborative design---creates persistent knowledge objects and
``paying it forward'' mechanisms that enable longitudinal community benefit
through shared resources \cite{pandya2025transformative,butcher2025community}.
This cross-boundary exchange enacts the co-dissemination phase of
transdisciplinary learning environments \cite{gulikers2025transdisciplinary}.

\subsection{The Bilingual Cultural Map Interface as Boundary Object}
\label{app:boundary}

Extending Sperling et al.'s analysis of hidden labor in educational AI
co-design \cite{Sperling2024}, the Bilingual Cultural Map Interface mediates
between university-based knowledge production, community expertise, and
public engagement without requiring semantic consensus. The platform
accommodates competing priorities---algorithmic efficiency for student
developers, narrative coherence for museum staff, cultural authenticity for
community members---sustaining cross-sector partnerships essential to SDG
implementation while distinguishing between university pedagogical authority
and community partner expertise.

\subsection{Full RQ3 Analysis: Emergent Properties of Collective Intelligence}
\label{app:rq3-full}

This section presents the complete analysis of RQ3 that was abbreviated in
the main text due to space constraints.

\textbf{Interdisciplinary Knowledge Synthesis.} The AI-enabled Community-Based
Learning instantiation advances from transactional ``helpers and recipients''
models toward reciprocal knowledge partnerships \cite{yahaya2025mapping},
addressing the critical gap where computation and design training privileges
problem-solving over ethical reflection \cite{lozano2026active}. The synthesis
of computational engineering and social work methodologies provided the
conceptual framework for negotiating ethical dimensions absent in purely
technical curricula. By importing community-engagement practices from social
work into software development, students bridged these disciplinary
boundaries, acquiring the vocabulary to translate technical choices---open
data standards, accessibility APIs, bilingual interfaces---into tangible
commitments to human dignity and intercultural dialogue.

\textbf{Philanthropic-Academic-Community Triads.} At the Third Cross-Strait
Hong Kong and Macao Service-Learning Student Conference at The Hong Kong
Polytechnic University (2026), student co-authors presented their work to
student peers, social work professors, and philanthropic foundation
representatives, demonstrating how cross-boundary integration advances both
fields reciprocally in alignment with UNESCO frameworks for digitally-enabled
community-based learning in higher education \cite{butcher2025community}.
For computational education, importing Community-Based Learning methodologies
addresses the critical gap where CS active learning privileges practical
problem-solving over ethical reflection---instead positioning technical work
as sustainable civic infrastructure that outlasts semester timelines
\cite{lozano2026active}. Reciprocally, for service-learning scholarship,
which documents persistent sustainability challenges and short funding
cycles that truncate community impact \cite{yahaya2025mapping,ridwan2025beyond},
the AI-enabled digital technology creates persistent knowledge objects and
``paying it forward'' mechanisms that enable longitudinal community benefit
through shared resources \cite{pandya2025transformative,butcher2025community}.
This cross-boundary exchange enacts the co-dissemination phase of
transdisciplinary learning environments \cite{gulikers2025transdisciplinary},
validating that reciprocal knowledge partnerships between computational and
social work methodologies dissolve disciplinary silos while addressing
sectoral gaps in both CS education and Community-Based Learning practice.

\textbf{Distributed Cognition Across Sectors.} The Bilingual Cultural Map
Interface functions as a ``boundary object'' \cite{Star1989}, enabling
coordinated action while supporting divergent interpretations:
student developers view it as modular code architecture for technical
learning; museum educators interpret it as curriculum infrastructure
extending their mission; civil society end-users understand it as accessible
cultural heritage facilitating place-based learning. Extending Sperling
et al.'s analysis of hidden labor in educational AI co-design
\cite{Sperling2024}, this reveals how the artifact mediates between
university-based knowledge production, community expertise, and public
engagement without requiring semantic consensus. The platform accommodates
competing priorities---algorithmic efficiency for student developers,
narrative coherence for museum staff, cultural authenticity for community
members---sustaining cross-sector partnerships essential to SDG
implementation while distinguishing between university pedagogical authority
and community partner expertise.

\end{document}